\documentclass[Numbered, nature]{sn-jnl}

\usepackage{graphicx}
\usepackage{dcolumn}
\usepackage{bm}
\usepackage{mathtools}
\usepackage{mathrsfs, amsmath, wasysym}
\usepackage[mathscr]{eucal}
\usepackage{xfrac}
\usepackage{graphicx}
\usepackage{dcolumn}
\usepackage{bm}
\usepackage{color}
\usepackage{tabularx}
\usepackage{braket}
\usepackage{hyperref}
\hypersetup{colorlinks,linkcolor=blue,citecolor=blue,urlcolor=blue}
\usepackage[normalem]{ulem}
\usepackage[all]{hypcap}
\usepackage[dvipsnames,usenames,table]{xcolor}
\usepackage{chngcntr}
\usepackage{amsmath, amssymb}
\usepackage{chemformula}

\begin{document}

\title{Database of Tensorial Optical and Transport Properties of Materials From the Wannier Function Method}

\author*[1]{\fnm{Zhenyao} \sur{Fang}}\email{z.fang@northeastern.edu}
\author[1]{\fnm{Ting-Wei} \sur{Hsu}}\email{hsu.ting@northeastern.edu}
\author*[1]{\fnm{Qimin} \sur{Yan}}\email{q.yan@northeastern.edu}

\affil[1]{\orgdiv{Department of Physics}, \orgname{Northeastern University}, \city{Boston}, \postcode{02115}, \state{Massachusetts}, \country{United States}}

\abstract{
The discovery and design of functional materials for energy harvesting and electronic applications require accurate predictions of their optical and transport properties. While several existing databases contain the first-order optical properties and the electron transport properties calculated from high-throughput first-principles calculations, the amount of material entries is often limited and those functional properties are often reported in scalar form. Comprehensive databases for the tensorial properties still remain inadequate, which prevents from capturing the anisotropic effect in materials and the development of advanced machine learning models that incorporate the space group symmetry of materials. Therefore, in this work we present the largest-to-date database of tensorial optical properties (optical conductivity, shift current) and the database of tensorial transport properties (electrical conductivity, thermal conductivity, Seebeck coefficient, thermoelectric figure of merit zT) for 7301 materials, calculated from the Wannier function method. The quality of the Wannier functions were validated by the maximal spread of the Wannier functions and by the comparison with the band structures from first-principles calculations, ensuring the accuracy of the calculated properties. These results contribute to the systematic study the functional properties for diverse materials and can benefit future data-driven discovery of candidate materials for optoelectronic and thermoelectric applications. 
}

\maketitle

\section{Background \& Summary}
The combination of first-principles calculation methods and high-throughput workflow has demonstrated potential in systematically predicting complex material properties and constructing large-scale material databases such as the Materials Project~\cite{Jain13MaterialsProject}, JARVIS~\cite{Choudhary20Jarvis}, AFLOW~\cite{Curtarolo12AFLOW}, and C2DB~\cite{Gjerding21C2DB_1, Haastrup18C2DB_2}. Those databases significantly accelerate the discovery and screening of functional materials with desirable properties for catalysis~\cite{Yao19Database_catalysis, Montoya17Database_catalysis}, optoelectronics~\cite{Hara21Database_opto, Rosen22Database_opto}, and quantum defects~\cite{Ferrenti20Database_defects, Wines23Database_defects}.

While these existing databases provide well-organized material information, they mainly focused on fundamental properties such as the formation energy, band gap, and thermodynamic stability. Other functional properties that are also commonly used to access the performance of materials in device applications still remain limited. For example, the optical conductivity (the dielectric function) quantifies the linear relation between the induced electric current (polarization) to the shined light of an arbitrary frequency~\cite{Fox10OpticalProperty}, which can help to determine the absorption and reflectance of materials and also to reveal the exotic electronic structures in materials such as the topological nodal points and the chirality of materials in experiments~\cite{Xu20CoSi, Wang25Chiral}. Besides, the shift current response, which is a second-order optical effect, describes the photocurrent generated by the shift of wave functions in non-centrosymmetric materials~\cite{Young12ShiftCurrent_1, Tan16ShiftCurrent_2}. As a major component of the bulk photovoltaic effect, it provides an alternative approach to overcome the Shockley–Queisser limit in conventional photovoltaic devices such as the p-n junction~\cite{Sauer23SQLimit_1, Dai23SQLimit_2}. Furthermore, transport properties, such as the electrical conductivity, the thermal conductivity, and the Seebeck coefficient, can characterize the ability of materials to conduct electric and heat currents~\cite{Giuseppe2014SolidStatePhysics}; their combination produces the thermoelectric figure of merit (zT) that describes the maximum efficiency of materials to convert heat into electricity~\cite{Kim15ZTFigureOfMerit_1, Zhou21ZTFigureOfMerit_2, Snyder17ZTFigureOfMerit_3}. Therefore, having a database with the above mentioned optical and transport properties can greatly benefit identifying promising materials, including those that have not been well characterized in experiments, for advanced device applications. 

Furthermore, since 2018, machine learning models such as graph neural networks have been used to predict various properties of solid-state materials~\cite{Xie18CGCNN, Fung21GNN, Reiser22GNN, Fang24GNN, Fang25GNN}. However, those models have primarily focused on scalar properties such as the formation energy and the band gap~\cite{Xie18CGCNN, Fung21GNN, Reiser22GNN}, and only few were applied to predict spectral properties such as the optical conductivity (as a function of photon frequency)~\cite{Hung24GNNOpt} or the thermal conductivity (as a function of temperature), even though there already exists established model architecture for sequential target prediction in the machine learning field~\cite{Lin22SequentialPropertyML, Kaur19SequentialPropertyML}. This limitation arises from the lack of sufficient high-fidelity data on the spectral properties of materials, which were often required for training deep machine learning models.

While previous attempts were made to construct material databases with the above mentioned optical and transport properties, those databases either only contain a limited amount of data, with only entries for first-order optical properties such as the optical conductivity and the dielectric function~\cite{Jain13MaterialsProject, Petousis17OpticalDatabase, Zhao22OpticalDatabase, Trinquet24OpticalDatabase, Garrity21HamiltonianDatabase, Cai19OpticalDatabase, Hasan21OpticalDatabase, Anderson15OpticalDatabase, Choudhary18OpticalDatabase}, or are calculated with a relatively low computational accuracy due to the computational cost~\cite{Ricci17OpticalDatabase}. Moreover, those databases mostly report the properties in scalar form, such as the trace of the optical response tensor only. However, these scalar properties are mostly useful to isotropic materials. For anisotropic materials, the full response tensor for the optical and transport properties are more accurate to assess their performance in experimental setup, because the tensor components are closely related to the space group symmetry and can reflect the directional dependence of the physical responses to external fields~\cite{Kaur23Anisotropy, Barman21Anisotropy, Ibanez20Anisotropy, Saberi17Anisotropy}. Besides, recent advances in machine learning models which utilize the equivariance relation of materials have shown potential, in principle, for predicting tensorial properties~\cite{Batzner22EquivariantGNN}, but they were rarely studied in practice due to the lack of tensorial property data, including the second-rank (such as the optical conductivity, electrical conductivity, and thermal conductivity tensors) and third-rank tensors (such as the shift current response tensor). Therefore, the development of reliable database for those tensorial spectral properties is essential to the design and application of advanced machine learning models in materials science.

Based on these motivations, in this work we present a database of the tensorial optical and transport properties of 7301 materials calculated from the tight-binding Hamiltonian in the basis of maximally localized Wannier functions. Our high-throughput workflow can automatically choose the projection functions and the optimal energy windows to construct the Wannier functions by analyzing the chemical orbital character of the electronic states near the Fermi level. After excluding the entries whose Wannier functions are not well-localized, we calculate the optical properties as a function of photon energy up to 2.5~eV, including the optical conductivity and shift current response tensor, and the transport properties as a function of temperature which ranges from 100 to 1000~K, including the tensorial electrical conductivity, Seebeck coefficient, thermal conductivity (with only electron contributions), and figure of merit (zT). We applied this workflow to all nonmagnetic elemental and binary materials with less than 20~atomic sites and with band gap less than 1~eV from the Materials Project~\cite{Jain13MaterialsProject}, which results in 10142 material entries in our initial set. After excluding those with unconverged relaxation calculations, 9235 entries remain. By further comparing the band structures from first-principles calculations and Wannier tight-binding Hamiltonian and excluding those with mean absolute error (MAE) of the band energy difference greater than 0.05~eV, we finally arrive at 7301 material entries. The average MAE of materials in the final database is 0.01~eV, suggesting the high quality of the constructed Wannier functions and the reliability of the calculated optical and transport properties. The presented workflow and database serves as a foundation for future data-driven discovery of functional materials and the development of advanced machine learning models for predicting complex material properties.

\section{Methods}
The high-throughput workflow was built based on python packages atomate2~\cite{Ganose25atomate2} and jobflow~\cite{Rosen24Jobflow}. As shown in Figure 1, it mainly consists of three stages: (1) first-principles calculations; (2) automatic Wannierization; (3) optical and transport property calculations.

\begin{figure}[htb]
\centering
\includegraphics[width=\linewidth]{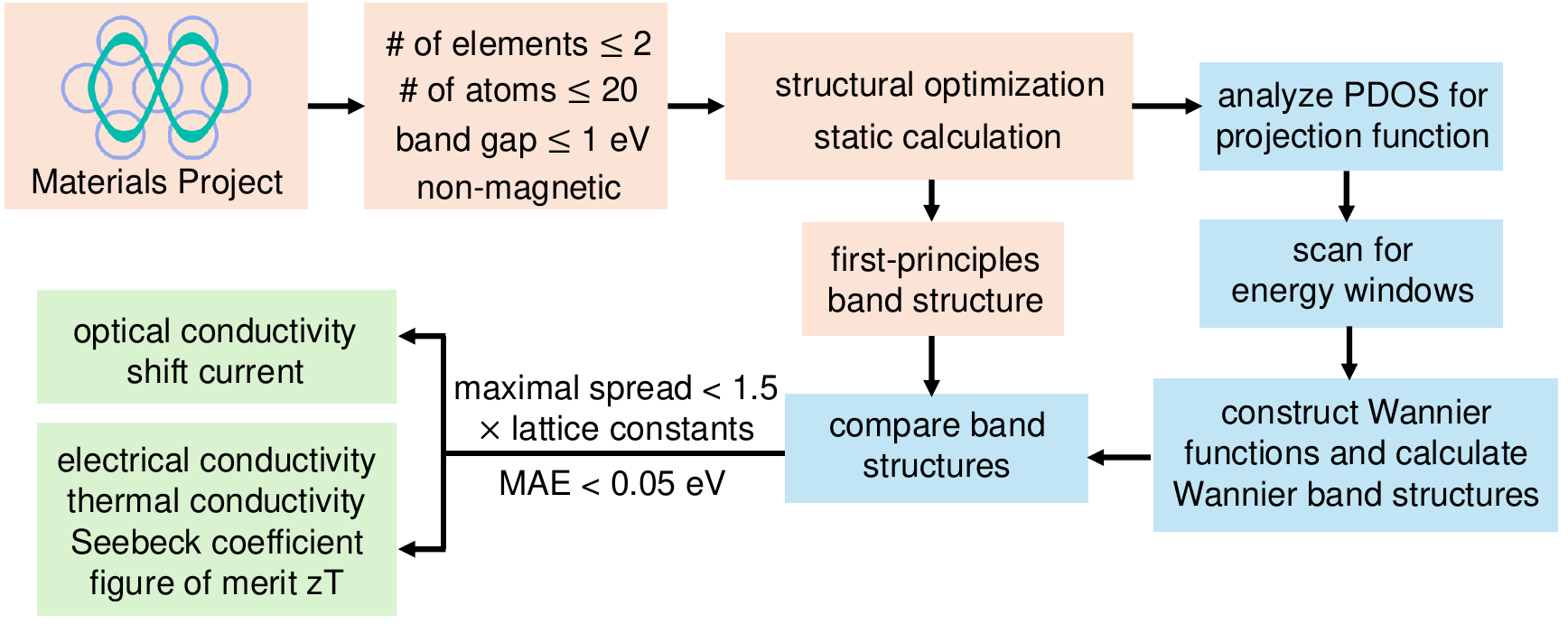}
\caption{A schematic plot of the high-throughput computational workflow to calculate tight-binding Hamiltonian and corresponding optical and transport properties. The pink, blue, and green blocks represent the first-principles calculation, automatic Wannierization, and optical and transport property calculation stages.}
\label{fig:workflow}
\end{figure}

\subsection{First-Principles Calculations}
To construct our database, we first select all elemental and binary nonmagnetic material entries from the Materials Project~\cite{Jain13MaterialsProject}. We excluded the materials with more than 20 atomic sites or with elements heavier than bismuth, and we also restricted to materials with band gap less than 1~eV. For each material entry, we first symmetrized the structure according to the procedure in Ref.~\cite{Setyawan10Symmetry} and performed several chained first-principles calculations to obtain the ground-state charge density and wave functions, as well as the electronic band structures along the high-symmetry lines for next-step comparison with the band structures calculated from Wannier tight-binding Hamiltonian. At this stage, material entries with unconverged structural relaxations will be excluded.

The first-principles calculations were performed using the Vienna Ab initio Simulation Package~\cite{Kresse93VASP1, Kresse96VASP2} with projector augmented wave pseudopotentials~\cite{Kresse99PAW1, Blochl94PAW2}. We used the Perdew-Burke-Ernzerhof functional in the generalized gradient approximation for all calculations~\cite{Perdew96GGA}. The kinetic energy cutoff for the plane-wave basis sets is 550~eV. The force convergence threshold is 0.01~eV/\AA, and the energy convergence threshold is $10^{-7}$~eV. For structural relaxation and self-consistent static calculations, we used the $\mathbf{k}$-point grid with the $\mathbf{k}$-point density of $0.03~(2\pi / \text{\AA})$, while for the non-self-consistent calculation we chose the $\mathbf{k}$-point density of $0.02~(2\pi / \text{\AA})$. All calculations were spin-unpolarized, since we only selected nonmagnetic materials, and also without the spin-orbit coupling effect. To take account of the strong correlation effect, we employed the DFT+U method~\cite{Anisimov91DFT_U, Cococcioni05DFT_U}, where the $U$ values are adapted from the high-throughput workflow of the Materials Project and summarized in Table 1~\cite{Jain13MaterialsProject, Jain11DFT_U}.

\begin{table}[htb]
    \centering
    \begin{tabular}{|c|c|c|c|c|c|c|c|c|}
    \hline
    & Co & Cr & Fe & Mn & Mo & Ni & V & W \\ \hline
    U (eV) & 3.32 & 3.7 & 5.3 & 3.9 & 4.38 & 6.2 & 3.25 & 6.2 \\ \hline
    \end{tabular}
    \caption{The $U$ values used for the DFT+U method.}
    \label{tab:DFT+U}
\end{table}

\subsection{Automatic Wannierization}
Maximally localized Wannier functions provide an efficient and physically intuitive representation of the electronic states~\cite{Marzari97WannierFunctions, Souza01WannierFunctions, Marzari12WannierFunctions}. Due to the localized nature of the Wannier functions, the tight-binding Hamiltonian in the basis of Wannier functions is sparse and can be efficiently converted into the Bloch Hamiltonian at an arbitrary $\mathbf{k}$-point through the Wannier interpolation formalism. Since calculating the optical and transport properties often requires the summation over a dense $\mathbf{k}$-point grid, the Wannier function method is more efficient than direct first-principles calculations. Furthermore, with Wannier functions, we can construct a minimal tight-binding Hamiltonian which only describe the bands near the Fermi level. This can significantly accelerate the the subsequent physical property calculations, since those properties typically involve electronic transitions between bands near the Fermi level with different electron occupation. Therefore, in this work, we chose to construct the tight-binding Hamiltonian in the basis of maximally localized Wannier functions and calculate subsequent physical properties, as implemented in Wannier90 package~\cite{Mostofi14Wannier90}.

Constructing high-quality Wannier functions requires the identification of the leading chemical orbitals as projection functions that contribute to the bands near the Fermi level and the proper choices of the disentanglement and frozen energy windows. Therefore, in the automatic Wannierization section, we aim at obtaining the optimal set of projection functions and energy windows to produce maximally localized Wannier functions.

Firstly, following the procedure in Ref.~\cite{Zhang18HighThroughputWannierization}, we analyze the total density of states from the static calculations and select an energy range $[E_\text{min}, E_\text{max}]$ which covers the energy range $[E_F - 2.5~\text{eV}, E_F + 2.5~\text{eV}]$, where $E_F$ is the Fermi energy. The value of 2.5~eV corresponds to the maximal photon energy in our optical property calculations (see next section), ensuring that the tight-binding Hamiltonian contain all states within this energy range and thus the accuracy of our calculated properties. Besides, we require that the bands within this chosen energy range $[E_\text{min}, E_\text{max}]$ are separated from other band manifolds, especially the core states and the unphysical states considerably above the Fermi level. Secondly, we analyze the projected density of states (PDOS) within this energy range $[E_\text{min}, E_\text{max}]$, and select the chemical orbitals with the largest relative contribution to the total density of states as the projection functions to construct the Wannier functions. Finally, to choose the optimal disentanglement and frozen energy windows, we scan the energy values within $[E_\text{min}, E_\text{max}]$ and select those which produce the minimal spread of Wannier functions. 

The quality of the constructed Wannier functions is mainly characterized by the maximum spread of the Wannier functions, which suggests the extent of localization of the Wannier functions~\cite{Marzari12WannierFunctions}. In this work, we require that the maximum spread is less than $1.5 \times \text{max}_{i} \{a_i \}$, where $a_i$ is the lattice constant along the Cartesian direction $i = x, y, z$. Besides, we also access the quality of the Wannier functions by comparing the band structures from first-principles calculations with those obtained from the Wannier function method, and we require that the MAE of the band energy difference is less than 0.05~eV. Only materials that satisfy both criteria are selected for subsequent optical and transport property calculations.

\subsection{Optical and Transport Property Calculations}
For the material entries which satisfy the above spread and MAE criteria, we calculate their tensorial optical and transport properties. The optical properties, including the optical conductivity and the shift current, are functions of the photon energy ranging from 0~eV to 2.5~eV; this energy range is generally adequate to detect and validate the exotic electronic structures using experimental techniques such as infrared spectroscopy or spectroscopic ellipsometer~\cite{Xu20CoSi}. The optical conductivity tensor $\sigma^{\alpha \beta} (\hbar \omega)$~\cite{Animalu67OpticalConductivity}, and the shift current tensor $\sigma^{\alpha \beta \gamma} (0; \hbar \omega, -\hbar \omega)$~\cite{Ibanez18ShiftCurrent} are calculated as 

\begin{align}
    &\sigma^{\alpha\beta} (\hbar \omega) = \frac{i e^2 \hbar}{(2 \pi)^3} \sum_{nm} \int  \frac{f_{m\mathbf{k}} - f_{n\mathbf{k}}}{E_{m\mathbf{k}} - E_{n\mathbf{k}}} \frac{v_{nm}^\alpha (\mathbf{k}) v_{mn}^\beta (\mathbf{k})}{E_{m\mathbf{k}} - E_{n\mathbf{k}} - (\hbar \omega + i \eta)} d\mathbf{k} \label{eqn: optical_conductivity}\\
    &\sigma^{\alpha \beta \gamma} (0; \hbar \omega, -\hbar \omega) = -\frac{i \pi e^3}{(2 \pi)^3 4 \hbar^2} \sum_{nm} \int (f_{n\mathbf{k}} - f_{m\mathbf{k}}) (I_{mn}^{\alpha\beta\gamma} + I_{mn}^{\alpha\gamma\beta}) \\ \nonumber
    &\qquad\qquad\qquad \times [\delta(E_{m\mathbf{k}} - E_{n\mathbf{k}} - \hbar \omega) + \delta(E_{n\mathbf{k}} - E_{m\mathbf{k}} - \hbar \omega)] d\mathbf{k} \label{eqn: shift_current}
\end{align}

Here $\alpha, \beta, \gamma$ are Cartesian directions, $m, n$ are band indices, $f_{m \mathbf{k}}$ and $E_{m \mathbf{k}}$ are the Fermi-Dirac occupation and the band energy, $v_{nm}^a$ is the velocity matrix element, and $I_{mn}^{\alpha\beta\gamma} = r_{mn}^\beta r_{nm}^{\gamma; \alpha}$, where $r_{mn}^\alpha$ is the position matrix element, and $r_{nm}^{\gamma; \alpha}$ is its generalized derivative to $\mathbf{k}^\alpha$. Finally, $\eta$ is the smearing parameter, which is set to be 0.05~eV; this value corresponds to the typical relaxation time of 10~fs in semiconductors and metals~\cite{Sernelius91RelaxationTime, Zhou20RelaxationTime}.

The transport properties, including the electrical conductivity, the thermal conductivity, the Seebeck coefficient, and the thermoelectric figure of merit (zT), are calculated as functions of temperature ranging from 100~K to 1000~K~\cite{Giuseppe2014SolidStatePhysics, Pizzi14TransportProperties}. Above this temperature range, those transport properties are strongly influenced by disordering effects, such as phase transitions, lattice vibrations, and configurational disorder effect, and may not be relevant for experimental validation. The transport function is defined as 
\begin{equation}
    \Sigma^{\alpha\beta} (E) = \frac{1}{(2\pi)^3} \sum_n \int v_n^\alpha (\mathbf{k}) v_n^\beta (\mathbf{k}) \delta(E - E_{n, \mathbf{k}}) \tau(n, \mathbf{k}) d\mathbf{k}
\end{equation}
where $\tau(n, \mathbf{k})$ is the electron relaxation time. In this work, we adopt the relaxation-time approximation and set $\tau = 10$~fs, which is the typical relaxation time in metals and semiconductors~\cite{Sernelius91RelaxationTime, Zhou20RelaxationTime}. From the transport function, the kinetic coefficient tensors at different orders $K_n^{\alpha\beta} (T)$ ($n = 0, 1, 2$) are defined as
\begin{equation}
    K_n^{\alpha\beta} (T) = \int (-\frac{\partial f(E, T)}{\partial E}) \Sigma^{\alpha \beta} (E) (E - E_F)^n dE
\end{equation}
The electrical conductivity, Seebeck coefficient, and the thermal conductivity (with only electron contributions) tensors are calculated respectively as
\begin{align}
    &\boldsymbol{\sigma} (T) = e^2 \mathbf{K}_0 (T) \\
    &\mathbf{S}(T) = \frac{1}{eT} \mathbf{K}_0^{-1} (T) \mathbf{K}_1 (T) \\
    &\boldsymbol{\kappa} (T) = \frac{1}{T} [\mathbf{K}_2 (T) - \mathbf{K}_0^{-1} (T) \mathbf{K}_1^2 (T)]
\end{align}
Finally, we calculate the thermoelectric figure of merit along the three Cartesian directions $\alpha = x, y, z$ only,
\begin{align}
    (\text{zT})_\alpha = \frac{S_{\alpha\alpha}^2 \sigma_{\alpha\alpha} T}{\kappa_{\alpha\alpha}}
\end{align}

The density of the $\mathbf{k}$-point grid for optical and transport properties is chosen to be $0.0015~(2\pi / \text{\AA})$ for materials with one atom in the unit cell and $0.003~(2\pi / \text{\AA})$ for others. The benchmark results over the $\mathbf{k}$-point density can be found in the Technical Validation section.

\section{Data Records}
The "entries.csv" file contains the mp-id of the material in the Materials Project~\cite{Jain13MaterialsProject}, the chemical formula, and the MAE of the band energy difference. Each folder, with name being the mp-id of the material, contains the optical conductivity (the real and imaginary parts), the shift current, the electrical conductivity, the Seebeck coefficient, the thermal conductivity, and the figure of merit zT, all in the tensorial form.

In the database, the unit of photon energy is eV, the temperature is K, the optical conductivity is $1 / (\Omega \cdot \text{m})$, the shift current is $\mu \text{A}/\text{V}^2$, the electrical conductivity is $1 / (\Omega \cdot \text{m})$, the Seebeck coefficient is $\text{V}/\text{K}$, the thermal conductivity is $\text{W} / (\text{m} \cdot \text{K})$, and the figure of merit zT is dimensionless. 

In Figure 2, we show the calculated optical conductivity, electrical conductivity, shift current of four material entries (GaAs, InN, AsPd\textsubscript{3}, BaAs\textsubscript{2}) in our database, where each line represents different tensor components. Note that in this work, we use the primitive cell, instead of the conventional cell, of each material to calculate the optical and transport properties. This could lead to different conventions on the definition of tensor components. For example, the $xyz$-component of the shift current of GaAs is negative between 0 and 2.5~eV in our work, while when using the conventional cell it is positive~\cite{Ibanez18ShiftCurrent}.

\begin{figure}[htb]
\centering
\includegraphics[width=\linewidth]{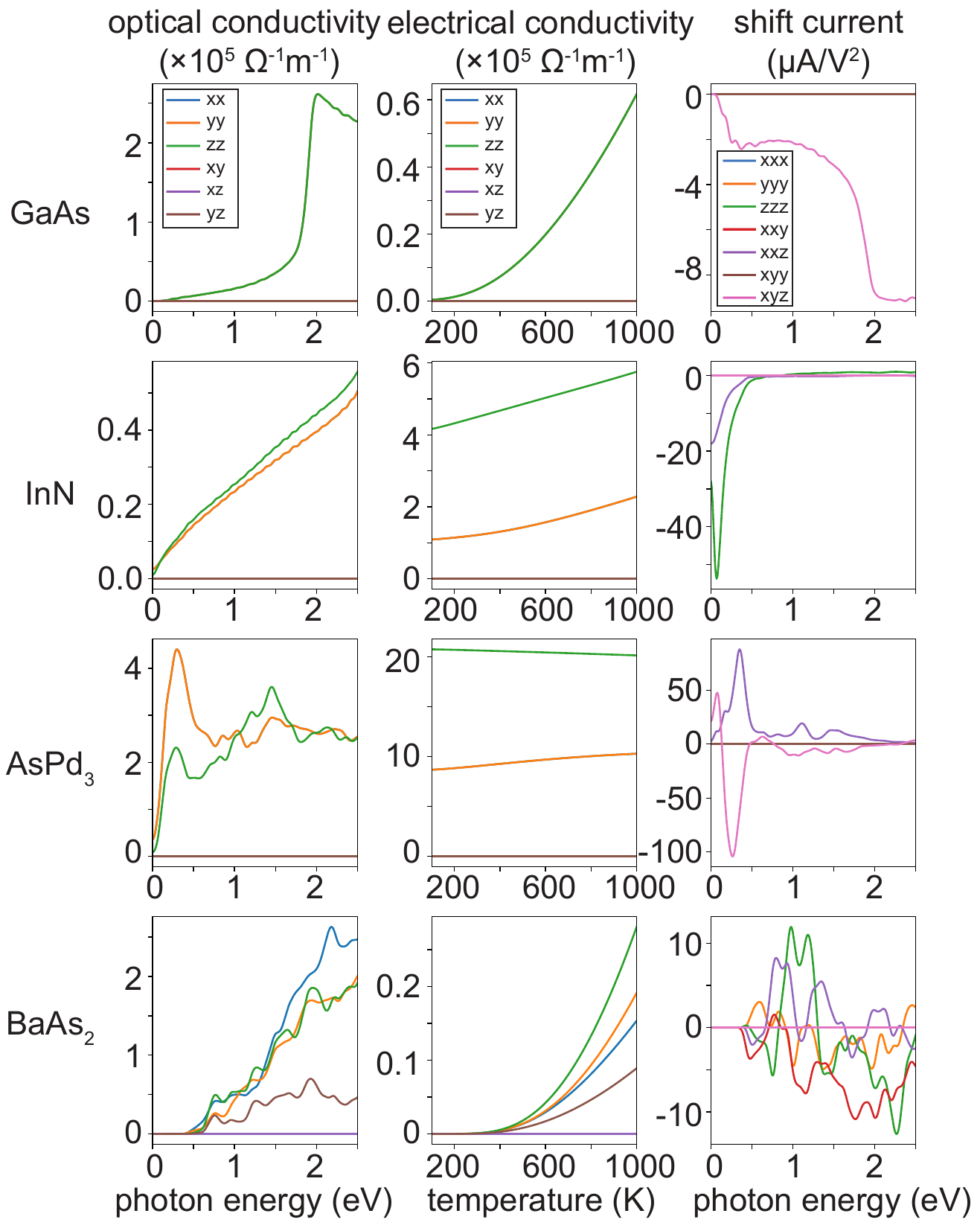}
\caption{Examples of the calculated optical conductivity, electrical conductivity, and shift current of GaAs, InN, AsPd\textsubscript{3}, BaAs\textsubscript{2}.}
\label{fig:sample_data}
\end{figure}

\section{Technical Validation}
The initial set of materials from the Materials Project, chosen according to the criteria in the Methods section, contains 10142 entries. After excluding those whose relaxation calculations cannot converge, this set restricts down to 9235 entries. Furthermore, by excluding materials whose Wannier functions do not satisfy the spread and MAE criteria, our final dataset contains 7301 entries.

As described in the Methods section, the maximal spread of the Wannier functions and the MAE of the band energy difference between the first-principles calculations and the Wannier function method are two important criteria to validate the quality of the Wannier functions. In Figure 3(a), we show the ratio of the maximal spread of the Wannier functions with respect to the maximum lattice constant for each material entry (whose relaxation calculations can converge), and 82.1\% of the material entries satisfy the criteria for the Wannier function spread (below the read dashed line). On the other hand, the MAE of the band energy differences is shown in Figure 3(b) and 80.5\% of the entries satisfy the criteria for the band energy difference (left to the red dashed line). After combining both criteria, we arrive at 7301 entries, which constitutes 79.1\% of the relaxed entries, for our final database.

\begin{figure}[htb]
\centering
\includegraphics[width=\linewidth]{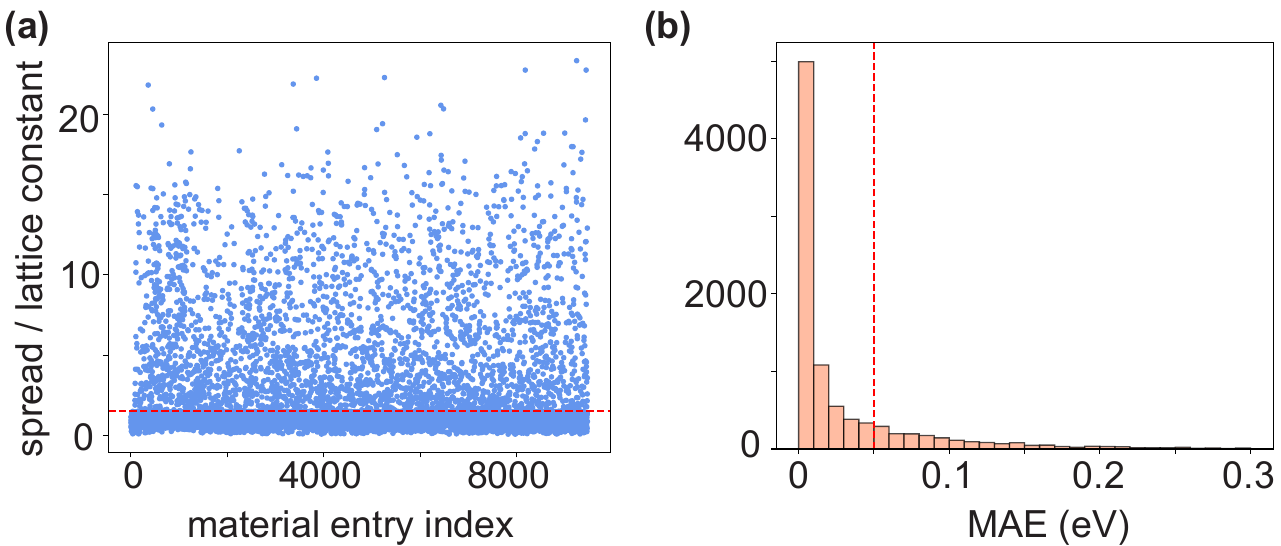}
\caption{(a) The ratio of the maximal spread to the maximum lattice constants for each material entry, and (b) the MAE of the band energy differences between the first-principles methods and the Wannier function method. The red lines represent our criteria for selecting materials for subsequent optical and transport property calculations.}
\label{fig:wannier_function_quality}
\end{figure}

Furthermore, we randomly selected 18 material entries with MAE smaller than 0.05~eV, and show the calculated band structures from first-principles calculations (black lines) and the Wannier function method (red lines) in Figures 4 and 5. In Figure 4, we show material entries where the MAEs are smaller than 0.02~eV; those material entries constitute 66.8\% of the 9235 materials whose relaxation calculations converged. The strong agreement between the two methods suggests the effectiveness and the universality of our high-throughput automatic Wannierization workflow. In Figure 5, we show the band structures material entries where the MAEs are between 0.02 and 0.05~eV. Although some inconsistencies are present, they usually occur at energies far above the Fermi level or at $\mathbf{k}$-points localized in the Brillouin zone; both cases do not affect the optical and transport properties as they all involve the electronic transitions between the bands close to the Fermi level and also the summation over a dense $\mathbf{k}$-point grid. These results suggest the validity of our criteria based on the MAE of band energy differences.

\begin{figure}[htb]
\centering
\includegraphics[width=\linewidth]{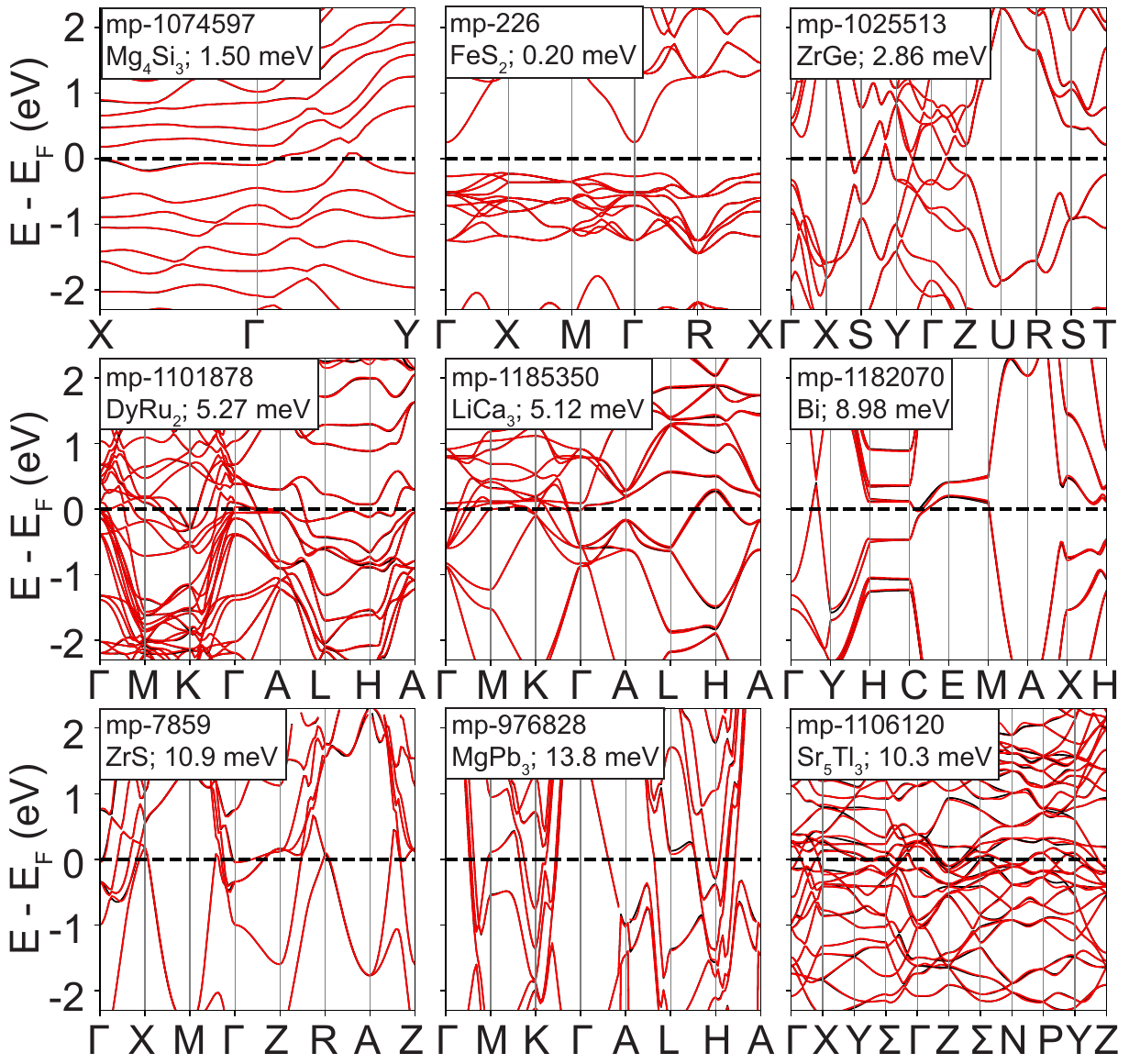}
\caption{The comparison of the band structures between the first-principles methods and the Wannier function method, where the MAEs of the band energy differences are smaller than 0.02~eV. The mp-id, chemical formula, and MAE for each entry are listed, and the black and red lines are band structures calculated from the first-principles and the Wannier function method, respectively.}
\label{fig:band_structure_I}
\end{figure}

\begin{figure}[htb]
\centering
\includegraphics[width=\linewidth]{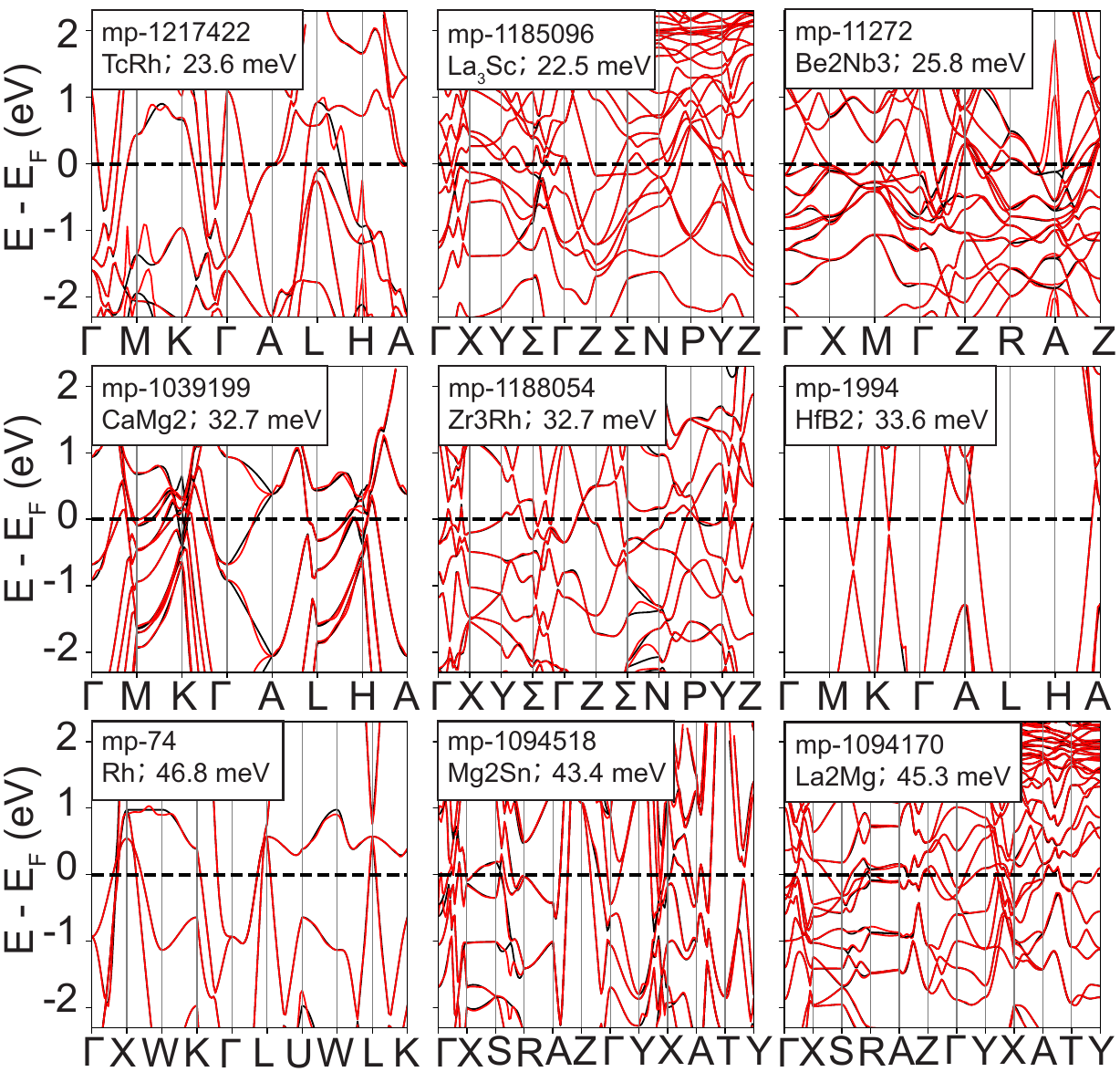}
\caption{The comparison of the band structures between the first-principles methods and the Wannier function method, where the MAEs of the band energy differences are between 0.02 and 0.05~eV. The mp-id, chemical formula, and MAE for each entry are listed, and the black and red lines are band structures calculated from the first-principles and the Wannier function method, respectively.}
\label{fig:band_structure_II}
\end{figure}

Having obtained the high-quality Wannier functions, we proceed to calculate the optical and transport properties. In Figure 6, we show the convergence test with respect to the $\mathbf{k}$-point grid. We chose four materials with different number of atoms in the unit cell: Al, GaAs, FeN, and BaAs\textsubscript{2}, with 1, 2, 2, 18 atoms in the unit cell respectively, respectively. For optical conductivity and electrical conductivity, we show the calculated $xx$-component. For shift current, due to the space group symmetry, we show the nonzero $xyz$-component for GaAs and the nonzero $xxx$-component for FeN and BaAs\textsubscript{2}. Note that the shift current response vanishes for the centrosymmetric crystal Al. Although we choose our $\mathbf{k}$-point grid for calculating the optical and transport properties by the $\mathbf{k}$-point density, we observe that for materials with small unit cells, a smaller density of 0.0015 ($2 \pi$/\AA) is still needed to obtain converged optical and transport properties, while for materials with larger unit cells, convergence is achieved under relatively coarse $\mathbf{k}$-point grid. Therefore, to construct our dataset, we chose the density of 0.0015 ($2 \pi$/\AA) for material entries with one atom in the unit cell, and 0.003 ($2 \pi$/\AA) for the others.

\begin{figure}[htb]
\centering
\includegraphics[width=\linewidth]{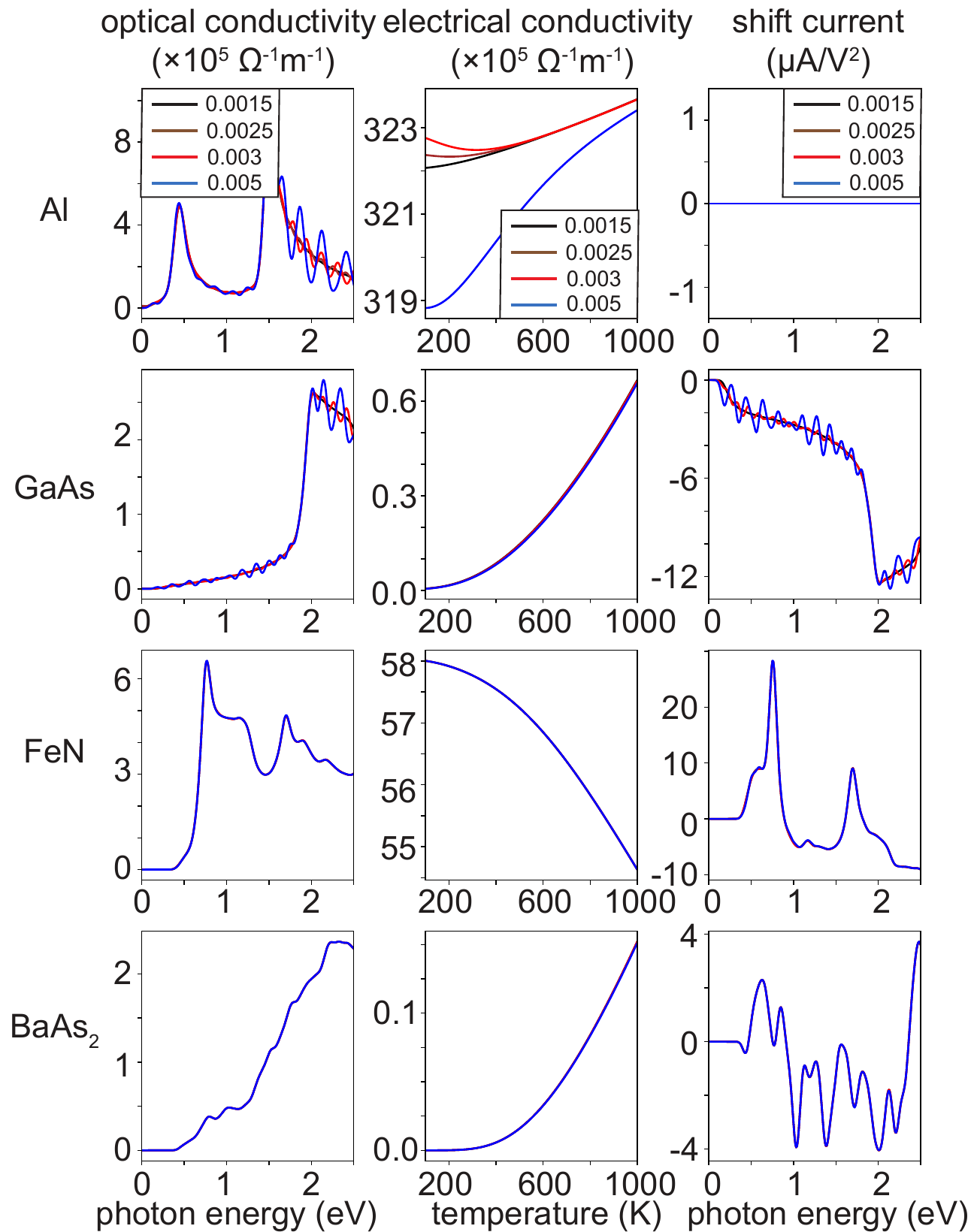}
\caption{The benchmark calculations over the $\mathbf{k}$-point grid density to calculate the optical and transport properties of Al, GaAs, FeN, and BaAs\textsubscript{2}, where the black, brown, red, blue lines represent the $\mathbf{k}$-point density of 0.0015, 0.0025, 0.003, 0.005~($2 \pi$/\AA).}
\label{fig:benchmark_kpoints}
\end{figure}

\section{Usage Notes}
The database in this work is the largest-to-date collection of systematically calculated optical and transport properties of materials using the Wannier function method. We are expanding the material space to multi-component and magnetic materials, and also using more advanced computational methods such as hybrid functions to correctly describe the electronic properties of materials. The user will be able to utilize the database to screen for functional materials with ideal optical and transport properties for experimental verification and machine learning model development. 

In this work, the relaxation time $\tau = 10$~fs was assumed to be a constant scaling factor to all transport properties; the user can scale the transport properties by other values of relaxation time, obtained from experiments or from electron-phonon coupling theory. Note that other work could report the electrical and thermal conductivity divided by the relaxation time~\cite{Ricci17OpticalDatabase} (the Seebeck coefficient and the figure of merit do not depend on $\tau$). Besides, when calculating the figure of merit zT, we didn't consider the lattice contribution to the thermal conductivity, so the reported figure of merit zT is in general overestimated.

\section{Code Availability}
The code used in this work is based on open-source python packages atomate2/0.0.17~\cite{Ganose25atomate2} and jobflow/0.1.18~\cite{Rosen24Jobflow}. The first principles calculations are performed using the Vienna Ab initio Simulation Package/6.4.2~\cite{Kresse93VASP1, Kresse96VASP2}, and the Wannierization is performed using the Wannier90 package/3.1.0~\cite{Mostofi14Wannier90}.

\bibliography{bibliography}
\bibliographystyle{naturemag}

\section{Author Contributions}
Z.F. and Q.Y. conceived the experiments. Z.F. wrote the code for the computational workflow and conducted the calculations. T.-W. H. helped with data analysis. Q.Y. supervised the study. The manuscript was written through contributions from all authors, and all authors have given approval to the final version of the manuscript.

\section{Competing Interests}
The authors declare no competing interests.

\section{Acknowledgments}
This work is supported by by the U.S. Department of Energy, Office of Science, Basic Energy Sciences, under Award No. DE-SC0023664. This research used resources of the National Energy Research Scientific Computing Center (NERSC), a U.S. Department of Energy Office of Science User Facility located at Lawrence Berkeley National Laboratory, operated under Contract No. DE-AC02-05CH11231 using NERSC award BES-ERCAP0029544.

\end{document}